\documentclass[aps,pre,twocolumn]{revtex4-2}

\usepackage{graphicx}
\usepackage{amsmath}
\usepackage{bm}
\usepackage{xcolor}

\begin{document}

\title{Phase Transitions of Polar Nematic Liquid Crystals Driven by Electrostatics and Flexoelectricity}
\author{Lincoln Paik}
\author{Jonathan V. Selinger}
\affiliation{Department of Physics, Advanced Materials and Liquid Crystal Institute, Kent State University, Kent, Ohio 44242, USA}
\date{\today}

\begin{abstract}
Polar liquid crystals can exhibit a series of phases and phase transitions, from the conventional nematic phase at high temperature to a modulated nematic (antiferroelectric) phase at intermediate temperature, and then to a uniform polar (ferroelectric) phase at lower temperature. Here, we develop a theory for this phase behavior. The theory is based on a competition between flexoelectricity (which favors splay in the modulated phase) and electrostatics (which provides an energy penalty for splay). Minimizing the free energy leads to differential equations that describe the spatial variation of the director, the polar order, and the electric potential. By solving the differential equations numerically, we determine the structure of the antiferroelectric phase. At high temperature, the modulated structure is sinusoidal. At lower temperature, it crosses over to a structure of wide domains separated by narrow domain walls. These numerical solutions can be compared with simplified models that apply near the phase transitions.
\end{abstract}

\maketitle

\section{Introduction}

An important advance in recent liquid-crystal research is the experimental discovery of nematic phases with polar order~\cite{Nishikawa2017,Chen2020,Lavrentovich2020}.  In a conventional nematic phase, molecules align in both directions along the director $\hat{\bm{n}}$.  By contrast, in a polar nematic phase, the symmetry between these two directions is broken, so that the material has a net electrostatic polarization.  Since their discovery, these phases have been extensively studied for basic research and for potential applications~\cite{Basnet2022,Sebastian2022,Mandle2022,Chen2022,Zavvou2022,Kumari2023,Perera2023,Zou2024,Basnet2025,Mathe2025}.

In any phase with electrostatic polarization, an important issue is how the polarization is structured as a function of position.  At high temperature, the polarization is zero and the system forms a conventional nematic ($N$) phase, which can be regarded as paraelectric.  At intermediate temperature, the system can form domains of alternating polarization, leading to an antiferroelectric ($N_{AF}$) phase, sometimes called a splay nematic or smectic-$Z_A$ phase~\cite{Mandle2017,Mertelj2018,Mandle2019,Sebastian2020,Connor2020,Chen2023,Ma2025,MedleRupnik2025,Ghimire2025,Badu2026}.  At low temperature, the polarization tends to be uniform over long length scales, and hence the material forms a ferroelectric nematic ($N_F$) phase.  In some cases, the uniform ferroelectric state can become unstable to the formation of a cholesteric helix~\cite{Khachaturyan1975,Kumari2024,Paik2025, Lavrentovich2025,Paik_2025(2)}.

Several experimental and theoretical studies have investigated the mechanism responsible for the intermediate antiferroelectric phase with alternating domains of polarization.  The earliest work by Mertelj and collaborators~\cite{Mertelj2018,Copic2020,Sebastian2020} identified the phase as splay nematic, and developed a model based on splay flexoelectricity~\cite{Dhakal2010,Shamid2013}.  Modeling by our group~\cite{Rosseto2020} confirmed that the one-dimensional (1D) splay modulation is a theoretical possibility, and also suggested a two-dimensional (2D) checkerboard modulation as an alternative possibility.  Emelyanenko \emph{et al.}~\cite{Emelyanenko2022} put this work into a general statistical theory of molecular ordering.  Chen \emph{et al.}~\cite{Chen2023} suggested an alternative model based on a free energy proposed by Shiba and Ishibashi~\cite{Shiba1978,Ishibashi1978}, which had previously been used for ferroelectric solids.

One limitation of those early models was that they did not explicitly include electrostatic interactions.  In a later paper~\cite{Paik2025}, we combined flexoelectricity and electrostatics, and showed that they have competing effects.  We considered electrostatic interactions associated with surface charges, and developed a model for the transition from conventional nematic to antiferroelectric.  A further study by Mertelj and collaborators~\cite{MedleRupnik2025} combined flexoelectricity and electrostatics through a different style of model, based on an assumed structure for the modulation, and modeled the evolution of the antiferroelectric structure over a wider temperature range.  Sprunt and collaborators~\cite{Badu2026} fit experimental data with a model based on the free energy of Shiba and Ishibashi~\cite{Shiba1978,Ishibashi1978}, with an approximation scheme using Jacobi elliptic functions~\cite{Klepikov1996}.

The purpose of this paper is to extend the theory of flexoelectricity and electrostatics in the intermediate antiferroelectric phase.  We go beyond our previous paper~\cite{Paik2025} by considering electric charges in the bulk rather than on the surface, and by modeling the full temperature range from conventional nematic through antiferroelectric, and into the uniform polar phase.  We go beyond Ref.~\cite{MedleRupnik2025} by allowing the modulation to take a more general form, and solving for the modulation that minimizes the free energy.  We find that the shape of the modulation changes with temperature.  It begins as a sinusoidal modulation near the nematic-antiferroelectric transition, and gradually evolves into a series of sharp domains and domain walls near the antiferroelectric-uniform transition.

The plan of this paper is as follows.  In Sec.~II, we develop the theory for a polar nematic liquid crystal with flexoelectricity and electrostatics, including electric charges in the bulk, and present numerical calculations in the intermediate antiferroelectric phase.  In Sec.~III, we investigate the nematic-antiferroelectric and antiferroelectric-ferroelectric phase transitions, and consider simplified versions of the theory that apply in those two cases.  In Sec.~IV, we discuss our results and compare them with previous theoretical research.

\section{Intermediate Antiferroelectric Phase}

\subsection{Model}

The intermediate antiferroelectric phase has both nematic order and polar order.  The nematic order can be described by the director field $\hat{\bm{n}}(\bm{r})$, with $\hat{\bm{n}}$ and $-\hat{\bm{n}}$ representing the same physical state, and the polar order can be described by the polarization field $\bm{P}(\bm{r})$.  We would like to develop a theory in terms of both of these fields.  We should not omit $\hat{\bm{n}}$, because there are places where $\bm{P}$ vanishes but nematic order is well-defined.  Likewise, we should not omit $\bm{P}$, because $\hat{\bm{n}}$ does not carry information about the magnitude or sign of polar order.

As in our previous paper~\cite{Paik2025}, we can write the total free energy for $\hat{\bm{n}}$ and $\bm{P}$ as the sum of four terms
\begin{equation}
F=F_\text{Frank} + F_\text{Landau} + F_\text{flexo} + F_\text{elec}.
\end{equation}
The first part is the Frank free energy
\begin{align}
	F_\text{Frank}&=\int d^3 r \biggl[\frac{1}{2}K_{11}(\bm{\nabla}\cdot\hat{\bm{n}})^2
	+\frac{1}{2}K_{22}(\hat{\bm{n}}\cdot\bm{\nabla}\times\hat{\bm{n}})^2\nonumber\\
	&\qquad\qquad+\frac{1}{2}K_{33}|\hat{\bm{n}}\times(\bm{\nabla}\times\hat{\bm{n}})|^2\biggr].
\end{align}
where $K_{11}$, $K_{22}$, and $K_{33}$ are the Frank elastic constants for splay, twist, and bend, respectively.

The second part is the Landau free energy as a power series in the polarization,
\begin{align}
	F_\text{Landau}&=\int d^3 r \biggl[\frac{1}{2}\mu |\bm{P}|^2 +\frac{1}{4}\nu|\bm{P}|^4
	+\frac{1}{2} \kappa|\bm{\nabla}\bm{P}|^2\biggr],
\end{align}
with $|\bm{\nabla}\bm{P}|^2=(\partial_i P_j)(\partial_i P_j)$.  As usual in Landau theory, the quadratic coefficient depends on temperature as $\mu(T)=\mu'(T-T_0)$, while the other coefficients are approximately independent of temperature.  In the quadratic term, we have dropped the distinction in our previous paper~\cite{Paik2025} between components of $\bm{P}$ parallel and perpendicular to $\hat{\bm{n}}$, for reasons that will be explained in Sec.~IV.

The third part is the flexoelectric coupling between polarization and splay,
\begin{equation}
	F_\text{flexo}=\int d^3 r \left[-\lambda(\bm{P}\cdot\hat{\bm{n}})(\bm{\nabla}\cdot\hat{\bm{n}})\right],
\end{equation}
as has been discussed in previous work~\cite{Rosseto2020,Selinger2022, Dhakal2010}.

The fourth part of the free energy is the electrostatic interaction energy, screened by ions dissolved in the liquid crystal.  We write it in the form 
\begin{equation}
	F_\text{elec}=\frac{1}{2}\int d^3 r \rho(\bm{r}) \phi(\bm{r}),
\end{equation}
where $\rho(\bm{r})=-\bm{\nabla}\cdot\bm{P}(\bm{r})$ is the electric charge density and $\phi(\bm{r})$ is the electrostatic potential.  The potential can be obtained by solving the screened Poisson equation
\begin{equation}
	\nabla^2 \phi(\bm{r}) = \frac{\phi(\bm{r})}{\Lambda_{D}^2} - \frac{\rho(\bm{r})}{\epsilon \epsilon_0}.
	\label{screenedpoissoneq}
\end{equation}
In this equation, $\Lambda_{D}$ is the Debye screening length given by
\begin{equation}
		\Lambda_{D}=\sqrt{\frac{\epsilon \epsilon_0 k_B T}{e^2 n}},
		\label{debye}
\end{equation}
where $\epsilon$ is the relative dielectric constant, $\epsilon_0$ is the permitivity of the vacuum, $k_B T$ is the thermal energy, and $n$ is the concentration of ions per volume in the material.  The limit of $\Lambda_{D}\to\infty$ corresponds to an unscreened electric interaction.  By comparison, in our previous paper~\cite{Paik2025}, we wrote this term in the alternative form
\begin{equation}
F_\text{elec}=\frac{1}{2}\int d^3 r_1 d^3 r_2 \frac{\rho(\bm{r}_1)\rho(\bm{r}_2) e^{-|\bm{r}_1-\bm{r}_2|/\Lambda_D}}{4\pi\epsilon\epsilon_0 |\bm{r}_1-\bm{r}_2|},
\end{equation}
which couples the electric charge density $\rho(\bm{r}_1)$ and $\rho(\bm{r}_2)$ at distinct positions $\bm{r}_1$ and $\bm{r}_2$.  These forms are equivalent provided that the potential $\phi(\bm{r})$ satisfies Eq.~(\ref{screenedpoissoneq}).
  
Combining all of the terms, we obtain the total free energy density
  \begin{align}
  	f &= \frac{1}{2}K{\large }_{11}(\bm{\nabla}\cdot\hat{\bm{n}})^2
  	+\frac{1}{2}K_{22}(\hat{\bm{n}}\cdot\bm{\nabla}\times\hat{\bm{n}})^2\nonumber\\
  	&\qquad+\frac{1}{2}K_{33}|\hat{\bm{n}}\times(\bm{\nabla}\times\hat{\bm{n}})|^2 + \frac{1}{2}\mu |\bm{P}|^2 +\frac{1}{4}\nu|\bm{P}|^4\nonumber\\
  	&\qquad +\frac{1}{2} \kappa |\bm{\nabla}\bm{P}|^2 -\lambda(\bm{P}\cdot\hat{\bm{n}})(\bm{\nabla}\cdot\hat{\bm{n}}) + \frac{1}{2} \rho(\bm{r}) \phi(\bm{r}).
\label{fcombined}
  \end{align}
We need to minimize the integral of this free energy density over the functions $\hat{\bm{n}}(\bm{r})$ and $\bm{P}(\bm{r})$, subject to the constraint of Eq.~(\ref{screenedpoissoneq}).

In order to explore the phase behavior numerically, it is helpful to reduce the number of free parameters in the model.  For that reason, we rescale the units of length, polarization, and energy to set $\lambda=1$, $K_{11}=1$, and $\kappa=1$.  In these units, the scaled free energy density takes the form
\begin{align}
	\tilde{f} &= \frac{1}{2} (\bm{\nabla}\cdot\hat{\bm{n}})^2
	+\frac{1}{2}\tilde{K}_{22}(\hat{\bm{n}}\cdot\bm{\nabla}\times\hat{\bm{n}})^2\nonumber\\
	&\qquad+\frac{1}{2}\tilde{K}_{33}|\hat{\bm{n}}\times(\bm{\nabla}\times\hat{\bm{n}})|^2 + \frac{1}{2} \tilde{\mu} |\bm{P}|^2 +\frac{1}{4}\tilde{\nu}|\bm{P}|^4\nonumber\\
	&\qquad +\frac{1}{2} |\bm{\nabla}\bm{P}|^2 -(\bm{P}\cdot\hat{\bm{n}})(\bm{\nabla}\cdot\hat{\bm{n}}) + \frac{1}{2} \rho(\bm{r}) \phi(\bm{r}),
	\label{scaledf}
\end{align}
and the screened Poisson equation constraint becomes
\begin{equation}
	\nabla^2 \phi(\bm{r}) = \frac{\phi(\bm{r})}{\tilde{\Lambda}_{D}^2} - \frac{\rho(\bm{r})}{\tilde{\epsilon}}.
	\label{scaledscreenedpoissoneq}
\end{equation}
In these expressions, the dimensionless coefficients are $\tilde{K}_{22}= K_{22}/K_{11}$, $\tilde{K}_{33}= K_{33}/K_{11}$, $\tilde{\mu}= \mu K_{11}/\lambda^2$,  $ \tilde{\nu}= \nu K_{11}^2/(\kappa \lambda^2)$, $\tilde{\epsilon} = \epsilon \epsilon_0 \lambda^2/ K_{11}$ and $\tilde {\Lambda}_D = \lambda  \Lambda_{D} / \sqrt{K_{11} \kappa}$.  Because the quadratic coefficient $\mu$ varies with temperature, the scaled parameter $\tilde{\mu}$ can be interpreted as a scaled temperature.  (For simplicity, we omit the tilde in the position $\bm{r}$, the gradient $\bm{\nabla}$, and the functions $\phi(\bm{r})$, $\bm{P}(\bm{r})$, and $\rho(\bm{r})$.)

As a next step, we assume that the director and scaled polarization fields have the one-dimensional modulations
\begin{subequations}
\label{ansatzgeneral}
\begin{align}
	\hat{\bm{n}}(x)&=(\sin\theta(x),0,\cos\theta(x)),\\
	\bm{P}(x)&= P(x) \hat{\bm{n}}(x) = P(x) (\sin\theta(x),0,\cos\theta(x)),
\end{align}
\end{subequations}
where $\theta(x)$ and $P(x)$ are functions to be determined.  This assumption is similar to the ansatz in our previous paper~\cite{Paik2025}, but not exactly the same.  The differences will be discussed in Sec.~IV below.  With that assumption, the free energy density takes the form
\begin{align}
	\tilde{f} &= \frac{1}{2}(\cos^2{\theta}+\tilde{K}_{33}\sin^2{\theta})\theta{'}^2+\frac{1}{2}\tilde{\mu}P^2
	+\frac{1}{4}\tilde{\nu}P^4-P\theta{'}\cos{\theta}\nonumber\\
	&\quad+\frac{1}{2}(P{'}^2+P^2\theta{'}^2)
	-\frac{1}{2}\phi(P{'}\sin{\theta}+P\theta{'}\cos{\theta}),
\end{align}
and the constraint is
\begin{align}
		\phi {''} = \frac{1}{\tilde{\Lambda}_D^2} \phi(\bm{x}) - \frac{\rho(\bm{x})}{\tilde{\epsilon}}.
\end{align}
We must now minimize the free energy over the functions $\theta(x)$ and $P(x)$.  This variational procedure gives the two Euler-Lagrange equations, 
\begin{equation}
  0=\frac{\delta  \tilde{f} }{\delta\theta(x)},\qquad
  0=\frac{\delta  \tilde{f} }{\delta  P(x)}.
\end{equation}
We must solve these two equations along with the constraint of the screened Poisson equation.  Hence, the numerical problem is to solve the three coupled differential equations
\begin{subequations}
\label{diffeqs}
\begin{align}
	0&=\theta {''} (P^2 + \cos^2{\theta} + \tilde{K}_{33} \sin^2{\theta}) - \theta {'}^2 (1- \tilde{K}_{33}) \cos{\theta} \sin{\theta} \nonumber \\
	&\quad -P {'} (\cos{\theta} - 2 P \theta {'})  - \frac{1}{2} P \phi {'} \cos{\theta}, \\
	0&=P{''} -P(\tilde{\mu} + \theta{'}^2) - \tilde{\nu} P^3 + \theta{'} \cos{\theta} - \frac{1}{2} \phi{'} \sin{\theta}, \\
	 0&=\phi {''} -\frac{1}{\tilde{\Lambda}_D^2} \phi(\bm{x}) - \frac{1}{\tilde{\epsilon}} [P{'} \sin{\theta} + P \theta{'} \cos{\theta}],
\end{align}
\end{subequations}
for the three functions $\theta(x)$, $P(x)$, and $\phi(x)$.

One simple solution of this system of differential equations is the \emph{conventional nematic} phase with $\theta(x)=0$, $P(x)=0$, and $\phi(x)=0$.  It has constant director $\hat{\bm{n}}=\hat{\bm{z}}$ and zero polarization, and its free energy density is $\tilde{f}=0$.  Another simple solution is the \emph{uniform polar} or \emph{ferroelectric nematic} phase with $\theta(x)=0$, $P(x)=\sqrt{-\tilde{\mu}/\tilde{\nu}}$, and $\phi(x)=0$.  It has constant director $\hat{\bm{n}}=\hat{\bm{z}}$ and nonzero polarization, and its free energy density is $\tilde{f}=-\tilde{\mu}^2/(4\tilde{\nu})$ (defined for scaled temperature $\tilde{\mu}\leq0$).

In addition to those two uniform phases, the system may also have a nonuniform phase, which can be regarded as \emph{modulated nematic}, \emph{splay nematic}, or \emph{antiferroelectric}.  To study the phase behavior of the model and the structure of the modulated phase, we must solve the differential equations numerically.

\subsection{Numerical calculations}

\begin{figure*}
	\includegraphics[width=\columnwidth]{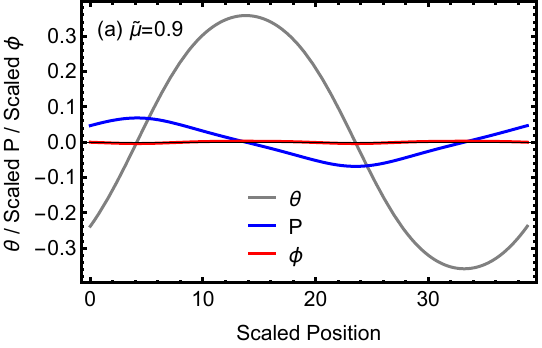}
	\includegraphics[width=\columnwidth]{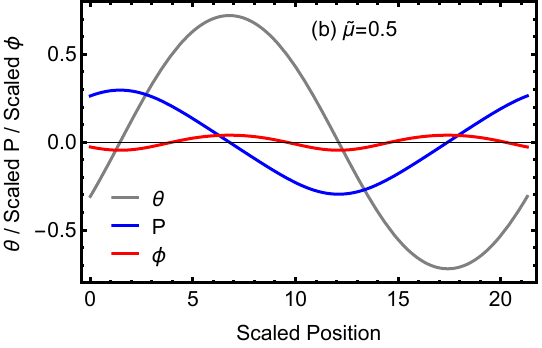}
	\includegraphics[width=\columnwidth]{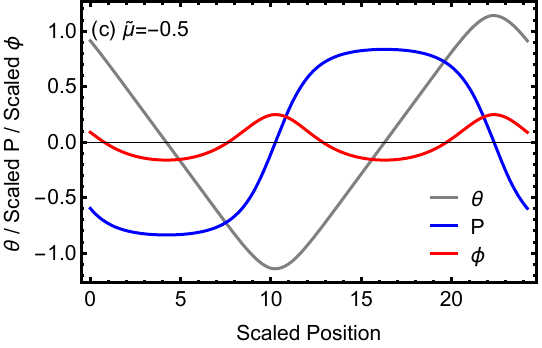} 
	\includegraphics[width=\columnwidth]{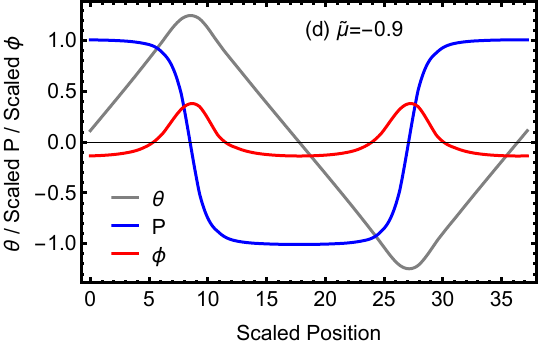}
	\includegraphics[width=\columnwidth]{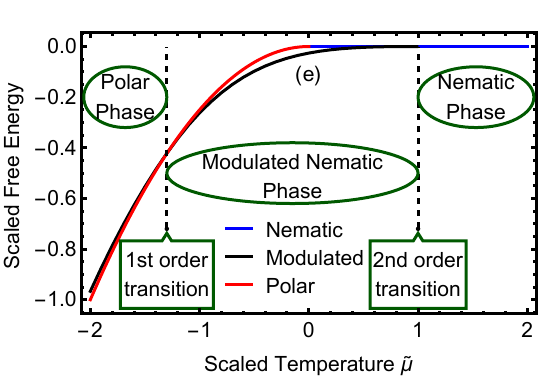}
	\includegraphics[width=\columnwidth]{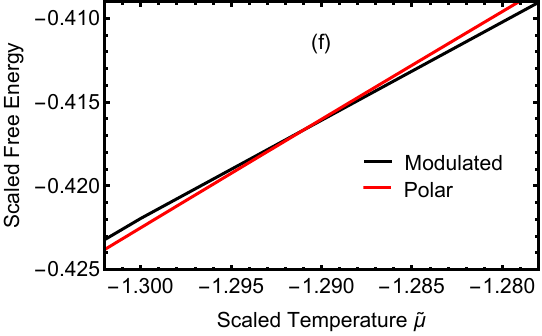}
	\caption{(a)--(d) Solutions of the set of differential equations~(\ref{diffeqs}) in the modulated nematic phase, at different values of the scaled temperature $\tilde{\mu}=0.9$, $0.5$, $-0.5$, and $-0.9$.  The other parameters are $\tilde{K}_{33} =1$, $\tilde{\nu} =1$, $\tilde{\epsilon} =1$, and $\tilde{\Lambda}_D =1$.  Note that the profiles cross over from sine waves to relatively sharp domains and domain walls as $\tilde{\mu}$ decreases.  (e) Scaled free energies of the nematic, modulated nematic, and uniform polar phases as functions of $\tilde{\mu}$, with labels indicating the phases and phase transitions.  (f) Zoomed-in view of the free energies of the modulated and uniform polar phases, showing the first-order transition.}
\end{figure*}

As a first step, we explore the effects of varying the scaled temperature parameter $\tilde{\mu}$, while keeping the other parameters fixed as $\tilde{K}_{33}=1$, $\tilde{\nu}=1$, $\tilde{\epsilon}=1$, and $\tilde{\Lambda}_D=1$.  At high temperature, for $\tilde{\mu} \geq 1$, the only numerical solution is the conventional nematic state with $\theta(x)=0$, $P(x)=0$, and $\phi(x)=0$.  As we reduce the scaled temperature below $\tilde{\mu}=1$, the numerical solution gives a modulated state with nonuniform $\theta(x)$, $P(x)$, and $\phi(x)$.  For each value of $\tilde{\mu}$, we solve the differential equations using periodic boundary conditions with a specified periodicity $\tilde{L}$, calculate the average free energy density, repeat for different $\tilde{L}$, and then choose the value of $\tilde{L}$ that minimizes the free energy.

Figure \hbox{1(a)--(d)} shows a series of numerical solutions for decreasing scaled temperature $\tilde{\mu}$, using the optimal $\tilde{L}$ in each case.  For $\tilde{\mu}=0.9$, just below the second-order transition, the functions $\theta(x)$, $P(x)$, and $\phi(x)$ all show sinusoidal modulations, and the wavelength $\tilde{L}$ is large.  At lower scaled temperature, for $\tilde{\mu}=0.5$, the modulations are still sinusoidal, the amplitudes become higher, and the wavelength becomes shorter.   As the scaled temperature becomes even lower, for $\tilde{\mu}=-0.5$ and $-0.9$, the modulations change into a different structure.  The polarization $P(x)$ now shows a series of positive and negative domains, separated by domain walls, which become sharper as $\tilde{\mu}$ decreases.  The director orientation $\theta(x)$ varies linearly in each domain, in the positive direction for one sign of $P(x)$ and the negative direction for the opposite sign.  The electric potential shows a peak in each domain wall.  The periodicity $\tilde{L}$ now increases as $\tilde{\mu}$ decreases.

Figure~1(e) shows a comparison of the average free energy densities for the three phases---conventional nematic (blue line), uniform polar (red line), and modulated nematic (black line).  At scaled temperature $\tilde{\mu}=0$, there is a \emph{second-order} transition from conventional nematic to modulated nematic.  At this transition, the derivative of free energy with respect to $\tilde{\mu}$ is continuous, and the order parameters vary continuously as functions of $\tilde{\mu}$.  At $\tilde{\mu}=-1.29$, there is a \emph{first-order} transition from modulated nematic to uniform polar.  At this transition, the free energies cross, and the state changes discontinuously as a function of $\tilde{\mu}$.  (The crossing of red and black lines can be seen more easily in the zoomed-in view of Fig.~1(f).)  Hence, we can identify the phases and phase transitions as labeled in Fig.~1(e).

We can now use the theory to model the effects of changing the ion concentration in the liquid crystal.  From Eq.~(\ref{debye}), we can see that the Debye screening length $\Lambda_D$ depends on ion concentration $n$ as $\Lambda_D\propto n^{-1/2}$.  Low ion concentration implies long screening length, so that electrostatic effects are long-ranged.  Conversely, high $n$ implies short $\Lambda_D$, so that electrostatic effects are highly screened and only short-ranged.  Hence, we solve the system of differential equations~(\ref{diffeqs}) for different values of the scaled screening length $\tilde{\Lambda}_D$, with all other parameters fixed at $\tilde{K}_{33}=1$, $\tilde{\nu}=1$, and $\tilde{\epsilon}=1$.

\begin{figure*}
	\includegraphics[width=\columnwidth]{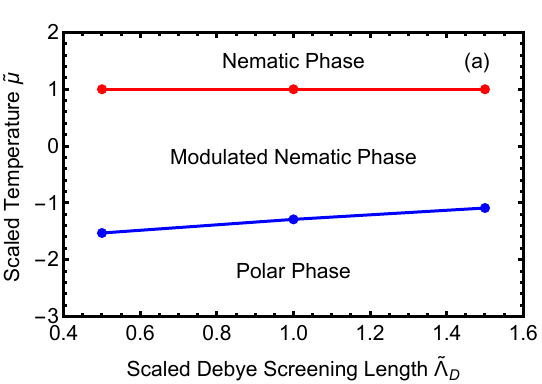}	
	\includegraphics[width=\columnwidth]{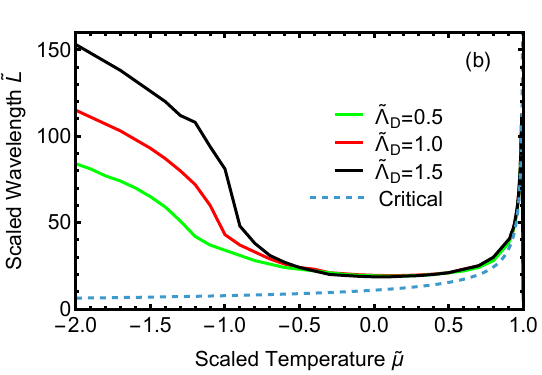}
	\includegraphics[width=\columnwidth]{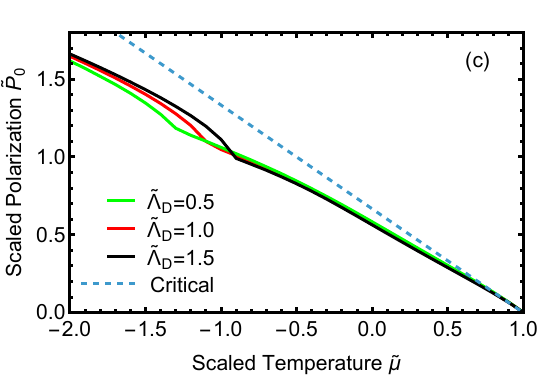}
	\includegraphics[width=\columnwidth]{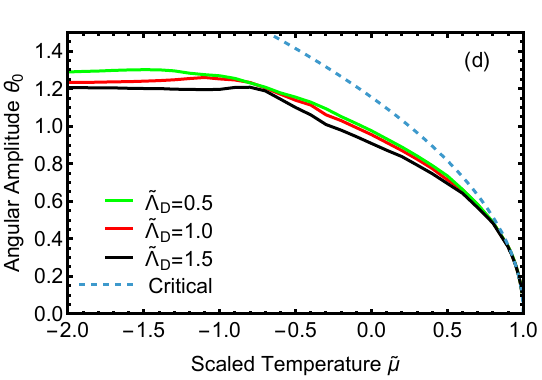}
	\caption{Effects of varying the scaled Debye screening length $\tilde{\Lambda}_D$ on the solutions of the differential equations (\ref{diffeqs}).  (a) Phase diagram.  (b)--(d) Plots of the scaled wavelength $\tilde{L}$, scaled polarization amplitude $\tilde{P}_0$, and angular amplitude $\theta_0$ as functions of scaled temperature $\tilde{\mu}$ through the modulated nematic phase.  The dashed lines show the critical behavior predicted by Eq.~(\ref{criticalpredictions}) in Sec.~III.}
\end{figure*}

The results are summarized in Fig.~2.  In the phase diagram of Fig.~2(a), there are conventional nematic, modulated nematic, and uniform polar phases.  The upper transition from conventional nematic to modulated nematic occurs at $\tilde{\mu}=1$, independent of $\tilde{\Lambda}_D$; it is apparently not related to electrostatic effects.  The lower transition from modulated nematic to uniform polar depends weakly on $\tilde{\Lambda}_D$, with higher screening length favoring the uniform polar phase and disfavoring the modulated phase.

Figure 2(b) shows the scaled wavelength $\tilde{L}$ as a function of scaled temperature $\tilde{\mu}$ for three values of $\tilde{\Lambda}_D$.  In all cases, $\tilde{L}$ is infinite at the second-order transition from conventional nematic to modulated.  As $\tilde{\mu}$ decreases through the modulated phase, $\tilde{L}$ first decreases and then increases again.  At low scaled temperature, $\tilde{L}$ depends on $\tilde{\Lambda}_D$, with a higher screening length giving a higher wavelength.  Figures 2(c) and 2(d) show the scaled polarization amplitude $\tilde{P}_0$ and angular amplitude $\theta_0$ as functions of $\tilde{\mu}$ for three values of $\tilde{\Lambda}_D$.  Both of these amplitudes depend sensitively on temperature but not on screening length.

Apart from the ion concentration, another way to vary the strength of electrostatic effects is through the dielectric constant $\epsilon$.  High values of $\epsilon$ correspond to weak electrostatic effects, and vice versa.  To model this dependence, we solve the system of differential equations~(\ref{diffeqs}) for different values of the scaled dielectric constant $\tilde{\epsilon}$.  For each choice of $\tilde{\epsilon}$, we adjust the scaled screening length as $\tilde{\Lambda}_D\propto(\tilde{\epsilon})^{1/2}$, because Eq.~(\ref{debye}) shows that these parameters are related in that way.  We keep the other parameters fixed at $\tilde{K}_{33}=1$ and $\tilde{\nu}=1$.

\begin{figure*}
	\centering
	\includegraphics[width=\columnwidth]{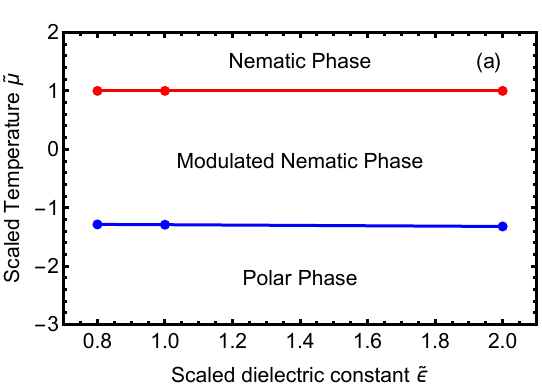}
	\includegraphics[width=\columnwidth]{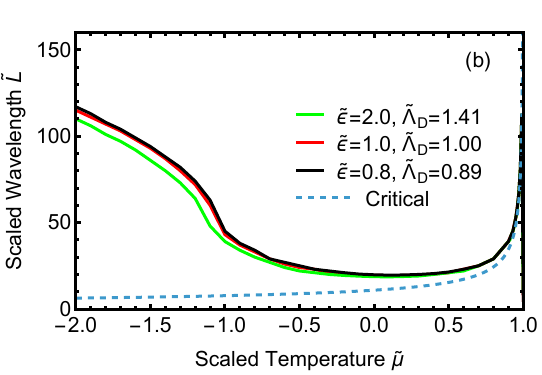}
	\includegraphics[width=\columnwidth]{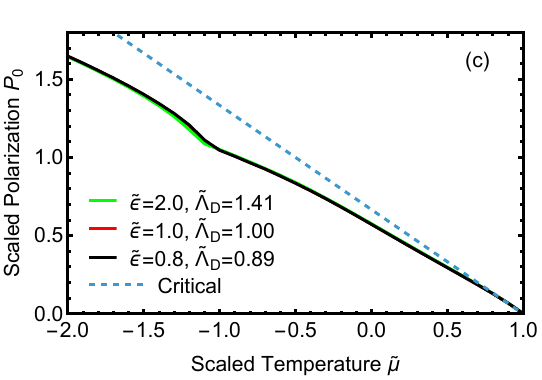} 
	\includegraphics[width=\columnwidth]{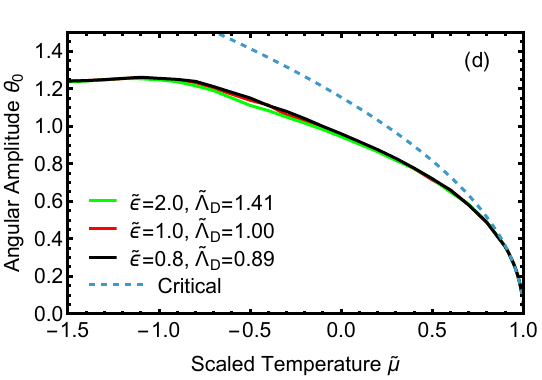}
	\caption{Effects of varying the scaled dielectric constant $\tilde{\epsilon}$, with a corresponding change in the scaled Debye screening length $\tilde{\Lambda}_D$, on the solutions of the differential equations~(\ref{diffeqs}).  (a) Phase diagram.  (b)--(d) Scaled wavelength $\tilde{L}$, scaled polarization amplitude $\tilde{P}_0$, and angular amplitude $\theta_0$ as functions of scaled temperature $\tilde{\mu}$ through the modulated nematic phase.  Dashed lines show the critical behavior predicted by Eq.~(\ref{criticalpredictions}).}
\end{figure*}

Figure~3 shows the effects of this variation of the dielectric constant.  In Fig.~3(a), we see the phase diagram for the conventional nematic, modulated nematic, and uniform polar phases.  The upper transition from conventional nematic to modulated nematic is at $\tilde{\mu}=1$, independent of $\tilde{\epsilon}$ and $\tilde{\Lambda}_D$.  The lower transition from modulated nematic to uniform polar is almost independent of the combined change in $\tilde{\epsilon}$ and $\tilde{\Lambda}_D$.  In Figs.~3(b)--(d), we see the wavelength, polarization amplitude, and tilt amplitude as functions of scaled temperature $\tilde{\mu}$ in the modulated nematic phase.  All of these results are almost independent of the combined change in $\tilde{\epsilon}$ and $\tilde{\Lambda}_D$.

Because Fig.~2 showed that changing $\tilde{\Lambda}_D$ by itself has some effects, we can infer that changing $\tilde{\epsilon}$ by itself has approximately the opposite effects.  To check this inference, we have also solved the differential equations while changing $\tilde{\epsilon}$ by itself, without changing $\tilde{\Lambda}_D$.  The results (not shown here) confirm that increasing $\tilde{\epsilon}$ by itself reduces the scaled wavelength $\tilde{L}$, while increasing $\tilde{\Lambda}_D$ increases $\tilde{L}$.  Neither $\tilde{\epsilon}$ nor $\tilde{\Lambda}_D$ has significant effects on the polarization amplitude or the tilt amplitude.

\section{Phase transitions}

In the previous section, we presented a numerical minimization of the free energy, which describes the full range of behavior from the conventional nematic phase, through the modulated nematic phase, and into the uniform polar phase.  For further insight, we can also develop approximations that apply near the phase transitions.  In this section, we consider first the upper transition from conventional nematic to modulated nematic, and then the lower transition from modulated nematic to uniform polar.

\subsection{Conventional to modulated nematic}

From the numerical minimization, we know that the system has a second-order transition from conventional nematic to modulated nematic.  On the modulated side of that transition, the system has sinusoidal variations of $\theta(x)$, $P(x)$, and $\phi(x)$, with small amplitudes and long wavelength.  To model this behavior, we make the variational ansatz $\theta(x)=\theta_0 \sin q x$ and $P(x)=P_0 \cos q x$.  From Eq.~(\ref{ansatzgeneral}), the director and polarization fields become
\begin{subequations}
\begin{align}
	&\hat{\bm{n}}(\bm{r})=(\sin(\theta_0\sin(q x)),0,\cos(\theta_0\sin(q x)),\\
	&\bm{P}(\bm{r})=P_0\cos(q x)(\sin(\theta_0\sin(q x)),0,\cos(\theta_0\sin(q x)).
\end{align}
\end{subequations}
With that ansatz, the electric charge density and electrostatic potential become
\begin{subequations}
\begin{align}
&\rho(\bm{r})=-\nabla\cdot\bm{P}(\bm{r}) = P_0 q \left[\theta_0\cos(2qx)+O(\theta_0^3)\right],\\
&\phi(\bm{r})=-\frac{P_0 q\Lambda_D^2}{\epsilon\epsilon_0}\left[\frac{\theta_0\cos(2qx)}{1+4q^2\Lambda_D^2}+O(\theta_0^3)\right],
\end{align}
\end{subequations}
as series in the small amplitude $\theta_0$.  We put these expressions into the free energy density of Eq.~(\ref{fcombined}) and average over position to obtain
\begin{align}
	f&=\frac{1}{8}K_{11}q^2\theta_0(\theta_0+J_{1}(2\theta_0))	+\frac{1}{8}K_{33}q^2\theta_0(\theta_0-J_{1}(2\theta_0))\nonumber\\
	&\quad+\frac{1}{4}\mu P_0^2+\frac{3}{32}\nu P_0^4+\frac{1}{16}\kappa P_0^2 q^2 (4+3\theta_0^2)\nonumber\\
	&\quad-\lambda P_0 q J_{1}(\theta_0)+\frac{P_0^2 q^2\theta_0^2\Lambda_{D}^2}{4\epsilon\epsilon_0(1+4q^2\Lambda_{D}^2)}.
\end{align}
First, we minimize over $\theta_0$, and obtain $\theta_0$ as a power series in $P_0$,
\begin{align}
	\theta_0&=\frac{\lambda P_0}{K_{11}q}+\frac{\lambda P_0^3}{8 K_{11}^4 q^3}\biggl[K_{11}\lambda^2-4K_{33}\lambda^2\\
	&\quad-2K_{11}^2 q^2 \left(3\kappa+\frac{4\Lambda_D^2}{\epsilon\epsilon_0(1+4q^2\Lambda_D^2)}\right)\biggr]+O(P_0^5).\nonumber
\end{align}
This minimization gives the effective free energy density in terms of $q$ and $P_0$,
\begin{align}
	f&=\frac{1}{4}\left[\mu-\frac{\lambda^2}{K_{11}}+\kappa q^2\right]P_0^2\nonumber\\ 
	&+\frac{1}{32}\biggl[3\nu+ \frac{6\kappa\lambda^2}{K_{11}^2}+\frac{2K_{33}\lambda^4}{K_{11}^4 q^2}
	+\frac{8\lambda^2\Lambda_D^2}{\epsilon\epsilon_0 K_{11}^2(1+4q^2\Lambda_D^2)}\biggr]P_0^4\nonumber\\
	&+O(P_0^6).
\end{align}
Next, we minimize over $q$, and obtain $q$ in power series in $P_0$,
\begin{equation}
q=\frac{\lambda K_{33}^{1/4}P_0^{1/2}}{2^{1/2}K_{11}\kappa^{1/4}}+O(P_0^{5/2}).
\end{equation}
That minimization gives the effective free energy density in terms of $P_0$ alone,
\begin{align}
f&=\frac{1}{4}\left[\mu-\frac{\lambda^2}{K_{11}}\right]P_0^2
+\frac{\lambda^2 K_{33}^{1/2}\kappa^{1/2}}{4K_{11}^2}|P_{0}|^3\\ 
&\quad+\frac{1}{32}\left[3\nu+\frac{2\lambda^2}{K_{11}^2}
\left(3\kappa+\frac{4\Lambda_D^2}{\epsilon\epsilon_0}\right)\right]P_{0}^4
+O(P_0^5).\nonumber
\end{align}
Now, we can minimize over $P_0$.  From the coefficient of quadratic term in the effective free energy, we see that there is a second-order transition from $P_0=0$ to $P_0\neq0$ when the temperature-like $\mu$ parameter reaches
\begin{equation}
	\mu_c= \frac{\lambda^2}{K_{11}}.
	\label{mucritical}
\end{equation}
Just below this transition, at $\mu=\mu_c-\delta\mu$, we have the critical behavior
\begin{subequations}
\begin{align}
P_0&=\frac{2K_{11}^2\delta\mu}{3\lambda^2 K_{33}^{1/2}\kappa^{1/2}}+O(\delta\mu^2),\\
q&=\frac{\delta\mu^{1/2}}{3^{1/2}\kappa^{1/2}}+O(\delta\mu^{3/2}),\\
\theta_0&=\frac{2K_{11}\delta\mu^{1/2}}{3^{1/2}\lambda K_{33}^{1/2}}+O(\delta\mu^{3/2}),\\
f&=-\frac{K_{11}^4\delta\mu^3}{27K_{33}\kappa\lambda^4}+O(\delta\mu^4).
\end{align}
\end{subequations}

If we rescale length, energy and polarization as in Sec.~II, then the critical temperature becomes simply $\tilde{\mu}_c=1$.  Just below the transition, at scaled temperature $\tilde{\mu}=\tilde{\mu}_c-\delta\tilde{\mu}$, the critical behavior is
\begin{subequations}
\label{criticalpredictions}
\begin{align}
\tilde{P}_0&=\frac{2\delta\tilde{\mu}}{3}+O(\delta\tilde{\mu}^2),\\
\tilde{q}&=\frac{\delta\tilde{\mu}^{1/2}}{3^{1/2}}+O(\delta{\tilde{\mu}}^{3/2}),\\
\theta_0&=\frac{2\delta\tilde{\mu}^{1/2}}{3^{1/2}}+O(\delta{\tilde{\mu}}^{3/2}),\\
\tilde{f}&=-\frac{\delta{\tilde{\mu}}^3}{27}+O(\delta{\tilde{\mu}}^4).
\end{align}
\end{subequations}
The predictions of Eq.~(\ref{criticalpredictions}) are shown by the dashed lines in Figs.~2-3.  We can see that they agree well with the numerical solutions of the differential equations near the upper transition from conventional nematic to modulated nematic.

Remarkably, the predictions of Eqs.~(\ref{mucritical})--(\ref{criticalpredictions}) do not depend on the electrostatic parameters $\Lambda_D$ and $\epsilon$.  Indeed, these predictions agree with Ref.~\cite{Rosseto2020}, which did not consider electrostatics.  From this lack of dependence, we can see that the upper transition from conventional nematic to modulated nematic is controlled by flexoelectricity, not by electrostatics.  Of course, electrostatic effects can be significant deeper in the modulated nematic phase, as shown by Fig.~2(b).  In Sec.~IV, we will present further discussion of electrostatics and the modulated phase.

\subsection{Modulated nematic to uniform polar}

We now consider the transition from the modulated nematic to the uniform polar phase.  From the solutions of the differential equations~(\ref{diffeqs}), shown in Fig.~1(c)--(d), we know that the functions $P(x)$ and $\theta(x)$ are not sine waves near this lower phase transition.  Rather, as the temperature decreases, these functions evolve toward a structure of domains separated by domain walls.  Within each domain, the polar order parameter $P(x)$ is approximately uniform, either positive or negative.  If $P(x)$ is positive, the orientation $\theta(x)$ increases approximately linearly, leading to positive splay $\bm{\nabla}\cdot\hat{\bm{n}}$.  If $P(x)$ is negative, $\theta(x)$ decreases approximately linearly, leading to negative splay $\bm{\nabla}\cdot\hat{\bm{n}}$.  These domains of approximately constant polarization and splay are separated by relatively narrow domain walls.  Within each wall, the orientation $\theta(x)$ is approximately constant, and $P(x)$ rapidly changes sign.

\begin{figure}
	\includegraphics[width=\columnwidth]{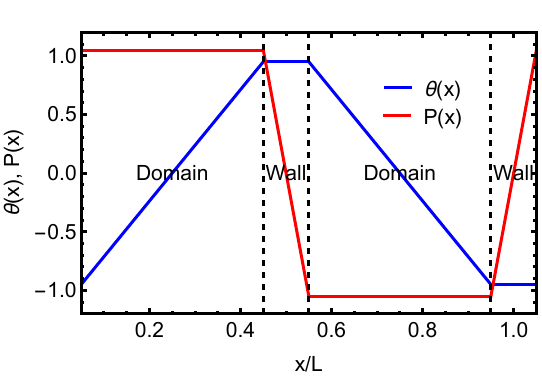}
	\caption{Simplified assumption for the polar order $P(x)$ and the orientation $\theta(x)$ in the modulated phase, near the transition to the uniform polar phase.  This model is used in the approximate theory of Sec.~III(B).}
\end{figure}

To model the behavior in the modulated phase near the lower transition, we assume that $P(x)$ and $\theta(x)$ have the simplified structure shown in Fig.~4, which is similar to the structure considered by Medle Rupnik \emph{et al}~\cite{MedleRupnik2025}.  In this simplified model, each domain has an exactly uniform value of $P$ and an exactly linear variation of $\theta$.  Likewise, each domain wall has a uniform value of $\theta$ and a linear variation of $P$.  Hence, the functions $P(x)$ and $\theta(x)$ are continuous, and their first derivatives are discontinuous.  This ansatz has four variational parameters:  the polarization amplitude $P_0$, the angular amplitude $\theta_0$, the overall periodicity $L$, and the width $d_w$ for each wall.  We assume that $d_w\ll L$, so that the walls are narrow in comparison with the domains.

We put the assumed form of $P(x)$ into the screened Poisson equation~(\ref{screenedpoissoneq}), and calculate the corresponding expression for the electric potential $\phi(x)$.  We then put the functions $P(x)$, $\phi(x)$, and $\theta(x)$ into the free energy density~(\ref{fcombined}), and calculate the total free energy in terms of the four variational parameters $P_0$, $\theta_0$, $L$, and $d_w$.  This calculation can be done exactly, but the results are too long to present here. We provide a Mathematica notebook in the Zenodo repository for any readers who would like to use those results.

We must now minimize the free energy over the four variational parameters.  That minimization can only be done numerically.  For the numerical work, we change to the scaled variables defined in Eqs.~(\ref{scaledf}--\ref{scaledscreenedpoissoneq}).  We choose specific values for the scaled temperature $\tilde{\mu}$ and the other scaled model parameters $\tilde{K}_{33}$, $\tilde{\nu}$, $\tilde{\epsilon}$, and $\tilde{\Lambda}_D$.  In each case, we minimize the free energy over the four scaled variational parameters $\tilde{P}_0$, $\theta_0$, $\tilde{L}$, and $\tilde{d}_w$.

\begin{figure*}
	\includegraphics[width=\columnwidth]{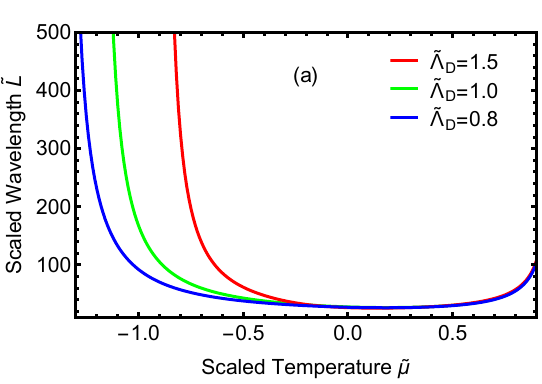}
	\includegraphics[width=\columnwidth]{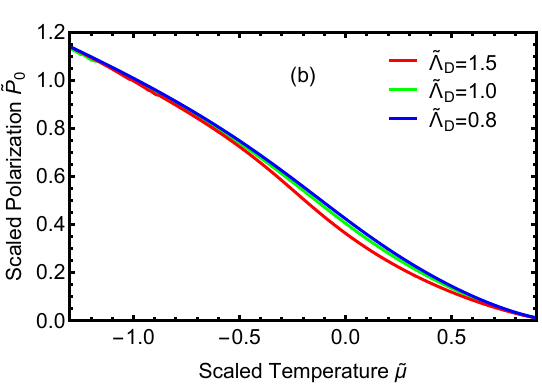}
	\includegraphics[width=\columnwidth]{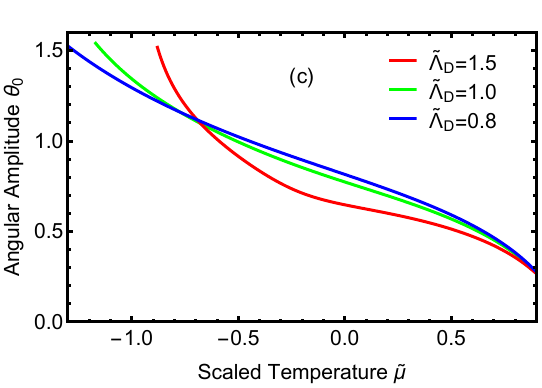}
	\includegraphics[width=\columnwidth]{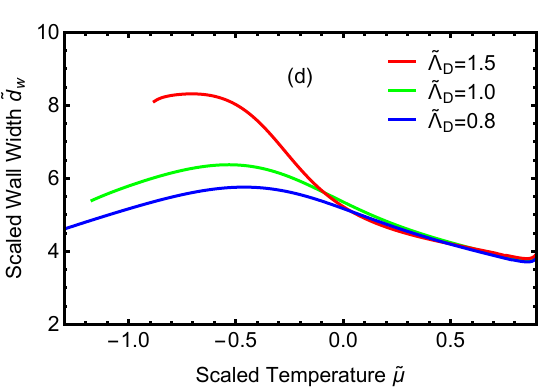}
	\caption{Numerical minimization of the free energy in the domain model of Sec.~III(B), showing the effects of varying the scaled temperature $\tilde{\mu}$ and the scaled Debye screening length $\tilde{\Lambda}_D$ for fixed $\tilde{K}_{33}=\tilde{\nu}=\tilde{\epsilon}=1$.  (a)~Scaled wavelength $\tilde{L}$.  (b)~Scaled polarization amplitude $\tilde{P}_0$.  (c)~Angular amplitude $\theta_0$.  (d) Scaled domain wall width $\tilde{d}_w$.}
\end{figure*}

\begin{figure*}
	\centering
	\includegraphics[width=\columnwidth]{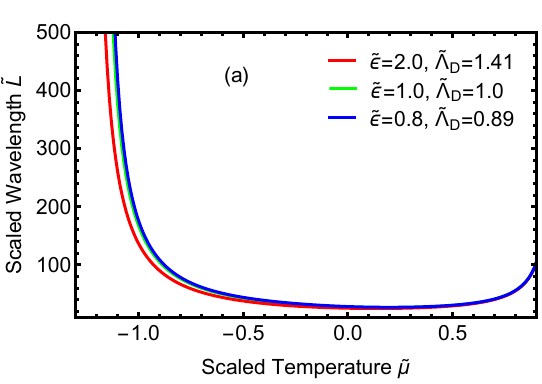}
	\includegraphics[width=\columnwidth]{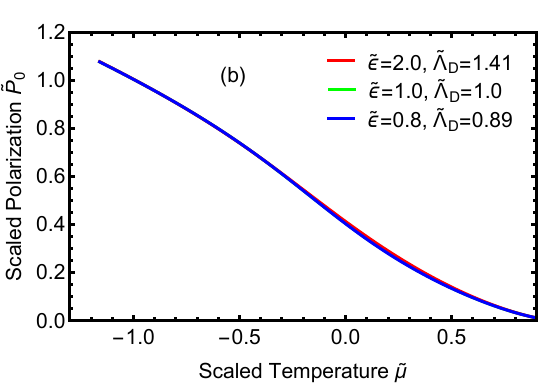}
	\includegraphics[width=\columnwidth]{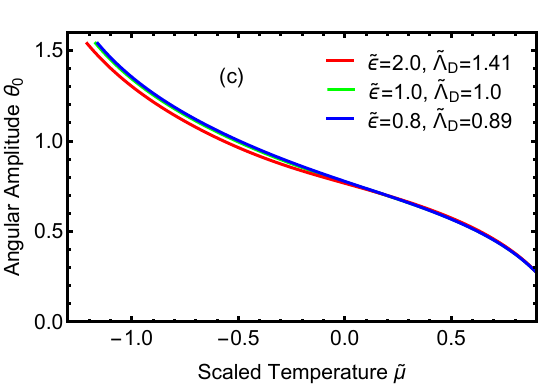}
	\includegraphics[width=\columnwidth]{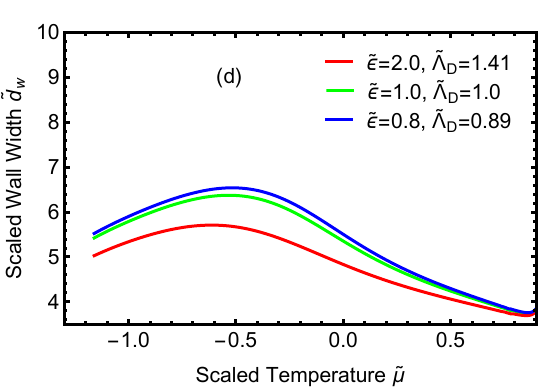} 
	\caption{Numerical minimization of the free energy in the domain model of Sec.~III(B), showing the effects of varying the scaled temperature $\tilde{\mu}$ and the scaled dielectric constant $\tilde{\epsilon}$, with a corresponding change in the scaled Debye screening length $\tilde{\Lambda}_D$, for fixed $\tilde{K}_{33}=\tilde{\nu}=1$.}
\end{figure*}

Results of the numerical minimization are shown in Figs.~5--6.  In each plot, the scaled temperature $\tilde{\mu}$ varies through the modulated nematic phase, from $\tilde{\mu}=1$ at the upper transition from the conventional nematic phase, to negative values at the lower transition to the uniform polar phase.  The scaled wavelength $\tilde{L}$ has a strong dependence on temperature.  It goes from a divergence at the upper transition to a short wavelength in the middle of the modulated nematic phase, and then increases again near the lower transition to the uniform polar phase.  The scaled polarization amplitude $\tilde{P}_0$ and angular amplitude $\theta_0$ both increase monotonically as the scaled temperature decreases from the upper to the lower transition.  The scaled domain wall width $\tilde{d}_w$ has a smaller variation.  The ratio $\tilde{d}_w/\tilde{L}$ becomes very small near the lower transition, which confirms that hypothesis of wide domains separated by narrow walls is self-consistent.

In Fig.~5, the different colors show the effects of varying the scaled Debye screening length $\tilde{\Lambda}_D$.  We can see that $\tilde{\Lambda}_D$ mainly affects the wavelength of the modulation.  As the screening length becomes shorter (i.e.\ the electrostatic interaction becomes more screened), the modulation wavelength $\tilde{L}$ becomes shorter, and the lower transition to the uniform polar phase shifts to lower temperature.  Also, as $\tilde{\Lambda}_D$ becomes shorter, the wall width $\tilde{d}_w$ becomes somewhat narrower.  The polarization amplitude $\tilde{P}_0$ and angular amplitude $\theta_0$ are not very sensitive to $\tilde{\Lambda}_D$.

In Fig.~6, the colors show the effects of varying the scaled dielectric constant $\tilde{\epsilon}$.  As in Fig.~3, for each choice of $\tilde{\epsilon}$, we adjust the scaled screening length as $\tilde{\Lambda}_D\propto(\tilde{\epsilon})^{1/2}$, because Eq.~(\ref{debye}) shows that these parameters are related in that way.  Here, we can see that this combined change has almost no effect on the results.  As $\tilde{\epsilon}$ decreases (so that the electrostatic interaction becomes weaker), $\tilde{\Lambda}_D$ also decreases (so that the electrostatic interaction becomes more screened).  These two changes approximately cancel each other in their effects on the modulation wavelength $\tilde{L}$ and wall width $\tilde{d}_w$.

\section{Discussion}

In the theory of polar liquid crystals, one general issue is:  What is the role of long-range electrostatic interactions?  Are they an essential part of the physics, or can the phase behavior be understood in terms of short-range elastic interactions?

The results presented in this paper suggest that the modulated phase is mainly controlled by short-range elastic interactions.  In our model, the second-order transition from the conventional to the modulated nematic phase is determined by the Frank, Landau, and flexoelectric parts of the free energy.  It is not affected by the electrostatic part of the free energy.  The behavior inside the modulated nematic phase is somewhat influenced by the electrostatic part of the free energy.  In particular, a stronger (less screened) electrostatic interaction causes the modulation wavelength to be larger.  However, the electrostatic interaction does not significantly influence the polarization and angular amplitudes in the modulated phase.  The transition from the modulated nematic to the uniform polar phase is also somewhat influenced by electrostatics.  A stronger (less screened) electrostatic interaction raises the temperature for this transition, favoring the uniform polar phase in comparison with the modulated nematic phase.  However, this transition still occurs even without electrostatics, driven by the increasing free energy cost of domain walls with decreasing temperature.

To the extent that electrostatic interactions are important, there are different ways for electrostatics to enter into the theory.  For that reason, we should emphasize a distinction between the model in our previous paper~\cite{Paik2025} and our current model.  In our previous paper, in the theory of the splay nematic phase, we made an ansatz for the polarization that was strictly along a single axis, perpendicular to the gradient direction.  The ansatz for $\bm{P}$ was approximately but not exactly parallel to the ansatz for $\hat{\bm{n}}$.  The angle between $\bm{P}$ and $\hat{\bm{n}}$ was limited by the coefficient $\mu_\perp$ in the free energy.  For that reason, the bound charge density was $\rho=-\bm{\nabla}\cdot\bm{P}=0$; i.e.\ there were no bound charges in the interior of the system.  Instead, we considered surface charges on the boundary of the system.  These surface charges had a significant effect on the transition from conventional to modulated nematic, but only if the system size $R$ was finite.  The effect went to zero in the thermodynamic limit of $R\to\infty$.

By contrast, in our current paper, the ansatz for $\bm{P}$ is strictly parallel to the director $\hat{\bm{n}}$.  Because these vectors are parallel, there is no need for the coefficient $\mu_\perp$.  Hence, the polarization has some splay, and the bound charge density $\rho=-\bm{\nabla}\cdot\bm{P}$ is nonzero in the interior of the system.  Our calculations use that bound charge density, not the surface charge density on the boundary.  As a result, the electrostatic effects do not involve the system size.  They persist even in the limit of an infinite system.

Overall, the model in this article provides a general view of the physics of the modulated nematic phase, and the effect of electrostatics on this phase.  It shows that the phase forms with a sinusoidal modulation of the director orientation and the polarization, and then crosses over to a structure of domains and domain walls.  At lower temperature, there is a transition to the uniform polar phase.  A stronger (less screened) electrostatic interaction causes the modulation wavelength to increase, and raises the transition temperature for the uniform polar phase.  Of course, the electrostatic interaction may have more profound effects on the behavior within the uniform polar phase.  These effects remain the subject of extensive ongoing research.

\section*{Acknowledgments}

We would like to thank S.~Sprunt for helpful discussions.

\section*{Conflicts of Interest}

The authors declare no conflicts of interest.

\section*{Data Availability Statement}

Details of our calculations are available in Mathematica notebooks in the Zenodo repository \url{https://zenodo.org/records/24139578}.

\bibliography{FlexoelectricNew}

\end{document}